\documentclass[sigconf,natbib=true]{acmart}
\usepackage{float}

\usepackage{algorithm}
\usepackage{algorithmic}
\usepackage{multirow}
\usepackage{subcaption}
\usepackage{multirow}

\begin{document}

\title{Modeling Temporal Positive and Negative Excitation for Sequential Recommendation}

\author{Chengkai Huang}
\authornote{Corresponding author.}
\email{chengkai.huang1@unsw.edu.au}
\affiliation{
  \institution{The University of New South Wales}
  \city{Sydney}
  \state{NSW}
  \country{Australia}
}

\author{Shoujin Wang}
\email{shoujin.wang@uts.edu.au}
\affiliation{
  \institution{University of Technology Sydney}
  \city{Sydney}
  \state{NSW}
  \country{Australia}
}

\author{Xianzhi Wang}
\email{XIANZHI.WANG@uts.edu.au}
\affiliation{
  \institution{University of Technology Sydney}
  \city{Sydney}
  \state{NSW}
  \country{Australia}
}

\author{Lina Yao}
\email{lina.yao@unsw.edu.au}
\affiliation{
  \institution{CSIRO’s Data 61 and UNSW}
  \city{Sydney}
  \state{NSW}
  \country{Australia}
}

\renewcommand{\shortauthors}{Chengkai, et al.}

\begin{abstract}

Sequential recommendation aims to predict the next item which interests users via modeling their interest in items over time. Most of the existing works on sequential recommendation model users' dynamic interest in specific items while overlooking users' static interest revealed by some static attribute information of items, e.g., category, or brand. Moreover, existing works often only consider the positive excitation of a user's historical interactions on his/her next choice on candidate items while ignoring the commonly existing negative excitation, resulting in insufficient modeling dynamic interest. The overlook of static interest and negative excitation will lead to incomplete interest modeling and thus impede the recommendation performance. To this end, in this paper, we propose modeling both static interest and negative excitation for dynamic interest to further improve the recommendation performance. Accordingly, we design a novel Static-Dynamic Interest Learning (SDIL) framework featured with a novel Temporal Positive and Negative Excitation Modeling (TPNE) module for accurate sequential recommendation. TPNE is specially designed for comprehensively modeling dynamic interest based on temporal positive and negative excitation learning. Extensive experiments on three real-world datasets show that SDIL can effectively capture both static and dynamic interest and outperforms state-of-the-art baselines.       

\end{abstract}



\maketitle

\section{Introduction}

Sequential recommender systems (SRSs) aim to generate sequential recommendations by predicting the next item which interests a given user. SRSs generally model the user's dynamic and timely interest in items from a sequence of historically interacted items for the recommendation. Benefiting from the strength of well-capturing users' interest changes over time, SRSs are able to provide more accurate and timely recommendation results to users~\cite{IJCAISurvey, wang2021survey}. 

\begin{figure}[t]  
	\centering
\includegraphics[width=\linewidth, height=4cm]{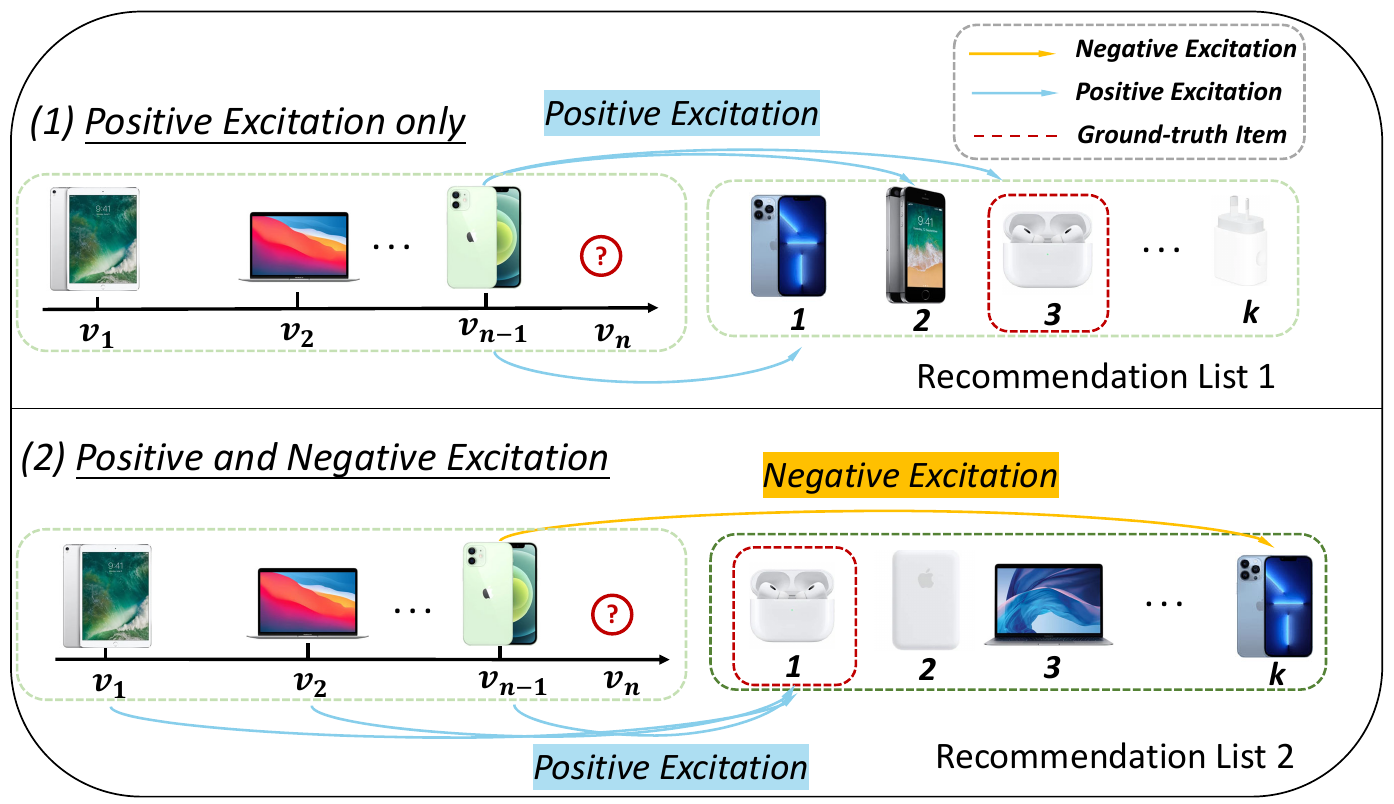}
	\caption{An example of recommendations via modeling positive excitation only (existing methods) and modeling both positive and negative excitation (our proposal). Clearly, the latter achieves better performance via ranking the ground-truth next item AirPods at the Top-1 position in the recommendation list.}
	\label{fig:TRDS}
  \Description{Figure 1. Fully described in the text.}
\end{figure}

A variety of SRS methods based on different models including Markov chain models, latent representation learning models, and deep learning models have been proposed in recent years. For instance, Markov chain-based SRSs adopt Markov chain models to model the first-order transitions over user-item interactions within an interaction sequence to predict the probable next item~\cite{shani2005mdp,FPMC}. Distributed representation learning-based methods map each interaction into a latent representation via capturing the contextual information for next-item recommendations~\cite{hu2017diversifying}. In recent years, advanced models including shallow or deep neural models have been employed in SRSs to improve the recommendation performance. For example, Recurrent Neural Networks (RNN) have been utilized to model the long- and short-term sequential dependencies in a sequence of interactions for next-item recommendations~\cite{GRU4Rec1}. Convolutional Neural Network (CNN)~\cite{Caser} and self-attention~\cite{SARS} models have been incorporated into SRSs for capturing more complex sequential dependencies (e.g., skip dependencies or collective dependencies) for next-item recommendations. These SRS methods have achieved remarkable recommendation performance.

Although remarkable recommendation performance has been achieved, there are still some significant gaps which prevent the further performance improvement of existing SRS methods. To be specific, \textit{most of the existing SRSs only model users' dynamic interest in items while overlooking users' static interest \cite{FDSA}} (\textbf{Gap 1}). However, static interest is also an important factor to determine which item to interact with for a given user. In this paper, \textbf{dynamic interest} means users' interest in specific items, which are captured from the ID information of a sequence of items interacted by each user. For instance, a user may be interested in iPhone 12. Item ID often changes frequently in most sequences, leading to frequent fluctuations of users' dynamic interest. \textbf{Static interest} refer to users' interest at a relatively higher level (e.g., item category or brand) and are usually more stable and change less frequently. For instance, a user may be interested in Apple products. Obviously, static interest changes less frequently than dynamic interest.   

More importantly, \textit{most of the existing SRSs cannot thoroughly 
capture users' dynamic interest since they often only model the positive excitation while overlooking negative one} (\textbf{Gap 2}). In SRSs, the \textbf{positive} (resp., \textbf{negative}) \textbf{excitation} refers to one prior interaction has a positive (resp., negative) impact on the next interaction. For instance, as an individual user, Alice may purchase a lens after she has purchased a Canon camera. However, it is unlikely that she will purchase another Nikon camera in a short period. In such a case, the purchase of Canon camera has positive excitation on the purchase of lens and negative excitation on that of Nikon camera. Clearly, negative excitation is an important signal to indicate which item may not be of the user's interest at the next time point and thus is necessary for accurate sequential recommendations. Although a few works in the literature have tried to model excitation for sequential recommendations, they generally only consider the positive excitation while ignoring the negative one ~\cite{SLRS}~\cite{AHMP}. For example, Wang et al. \cite{SLRS} defined a short-term and life-long positive self-excitation function for next-item recommendations.          

Aiming at bridging the aforementioned two significant gaps in exiting SRSs, we propose a novel \textbf{S}tatic-\textbf{D}ynamic \textbf{I}nterest \textbf{L}earning (\textbf{SDIL}) framework. SDIL is able to comprehensively model users' static and dynamic interest in items for generating accurate sequential recommendations. SDIL consists of (1) a Static Interest Modeling (SIM) module, (2) a Dynamic Interest Modeling (DIM) module, and (3) a next-item prediction module. To be specific, given a sequence of items interacted by a user, the corresponding attribute information of these items is input into the SIM module to capture the static interest of the user. Meanwhile, the ID information of these items is imported into the DIM module to comprehensively capture the user's dynamic interest. Then, both types of interest are well integrated as the input of the downstream prediction module for the next-item prediction. 

More importantly, we devise a novel \textbf{T}emporal \textbf{P}ositive and \textbf{N}egative \textbf{E}xcitation Modeling (TPNE) framework to quip the DIM module, which constitutes the main contribution of this work. Inspired by \cite{SLRS}, TPNE is built on the basis of the temporal point process (TPP)~\cite{TPP}, which is a classical model for modeling discrete event sequences in a continuous time period. In TPNE, first, a relation-based temporal module is designed to model the positive or negative excitation of each of a user's historical interactions on his/her next choice on candidate items. Here, the relation mainly refers to the substitute/complementary relations between items, which are extracted from the "co-click" (of items), "co-purchase" data. Then, a novel time decay kernel function is particularly designed to measure the excitation strength of each historical interaction according to the time interval between it and the next choice. Time interval is a key factor to determine whether the excitation strength is strong or weak.~\cite{SLRS,AHMP} 

The main contributions of this work can be summarized below:
\begin{itemize}

\item We propose a novel Static-Dynamic Interest Learning (SDIL) framework for comprehensively modeling users’ interest in items. SDIL consists of a static interest modeling (SIM) module, a dynamic interest modeling (DIM) module, and a next-item prediction module.  

\item We devise a novel Temporal Positive and Negative Excitation Modeling (TPNE) framework to quip the DIM module. TPNE is good at modng both the positive and negative excitation as well as the excitation strength.      
\item We evaluate our SDIL on three real-world datasets with different characteristics. Extensive experiments not only show the consistent superiority of SDIL over state-of-the-art baselines but also verify the rationality and effectiveness of our design in SDIL.  
\end{itemize}

\section{Related Work}
\subsection{Sequential Recommendation}

Sequential recommendation intends to recommend items that may interest users by modeling the sequential dependencies across users' historical interactions. A variety of methods have been proposed for the SRS task including rule-based methods~\cite{Yap}, KNN-based methods~\cite{JannachL17}, Factorization machine-based methods, and Markov chain-based methods~\cite{FPMC, FusingM}. 
Nevertheless, these methods fail to model the long user behavior sequences in terms of the limited representation ability of models.
Recent years have witnessed the success of deep learning-based methods applying in user modeling~\cite{VBPR, HeM16}, various models like CNN-based model~\cite{Caser}, RNN-based models~\cite{GRU4Rec1}~\cite{GRU4Rec2}, self-attention based models~\cite{SARS}~\cite{zhang2019next} were explored for SRS task. We briefly review attention-based SRS methods, which are most relevant to our work. 
The self-attention mechanism was proposed by Vaswani et al.~\cite{Transformer}. Benefiting from its strong ability to model the correlations among context information, many works employed it in SRS by intensifying the correlation between important historical items and target items. Li et al. proposed the NARM~\cite{NASR}, which leverages a global and local encoder with an attention module to model both user short- and long-term interest. Wang et al.~\cite{WangHCHL018} propose the ATEM, which is an attention-based transition embedding model. It builds the attentive context embedding without order assumption. Kang et al.~\cite{SARS} utilizes the transformer architecture to bridge the item-item relationship, which gains significant performance in the next-item prediction task.
However, all of these methods only focus on the item relations s within sequences, overlooking the vital temporal signals, which might lead to the sequential patterns being hard to generalize on unseen timestamps or different time intervals.
\subsection{Time-sensitive Recommendation}
Temporal recommendation considers the temporal evolution of user interest in item transition, represented by methods based on matrix factorization. 
TimeSVD++~\cite{TimeSVD} achieved strong results by splitting time into several bins of segments and combining it into the CF framework. Bayesian Probabilistic Tensor Factorization(BPTF) ~\cite{XiongCHSC10} is proposed to include time as an extra dimension for the tensor factorization, which is an extension of traditional MF methods. After that, a variety of works attempt to model the long- and short-term user interest separately. For instance, Zhu et. al.~\cite{TimeLSTM} propose the TimeLSTM to capture users' long and short-term dynamic interest through the specific time gates. Cai et. al. ~\cite{CaiBWSSW18} propose the LSHP, combining the short-term and long-term Hawkes Processes based on different segments of user history to predict the type and time of the next action in sequential online interactive behavior modeling. Wang et. al.~\cite{SLRS} also leverages the Hawkes Processes to model the user repeat consumption behavior from the perspective of short- and life-long terms. After that, Wang et al.~\cite{chorus} designs the multiple forms of temporal decay functions for item relations in the user interaction history. Recently, Wang et al.~\cite{KDA} have devised relational intensity and frequency domain embeddings to adaptively determine the importance of historical interactions. 
Besides, Fan et.al.~\cite{FanLZX0Y21} design a new framework TGSRec upon a pre-defined continuous-time bipartite graph, which can capture collaborative signals from both users and items, as well as consider temporal dynamics within sequential patterns. Although these time-sensitive methods well model the user's dynamic interest changes, they do not explicitly model the user's relatively stable interest. Meanwhile, most methods attempt to learn the positive temporal collaborative signals. but ignoring the effectiveness of negative temporal signals. 

\begin{figure*}[htp]
	\centering
	\includegraphics[width=1\linewidth]{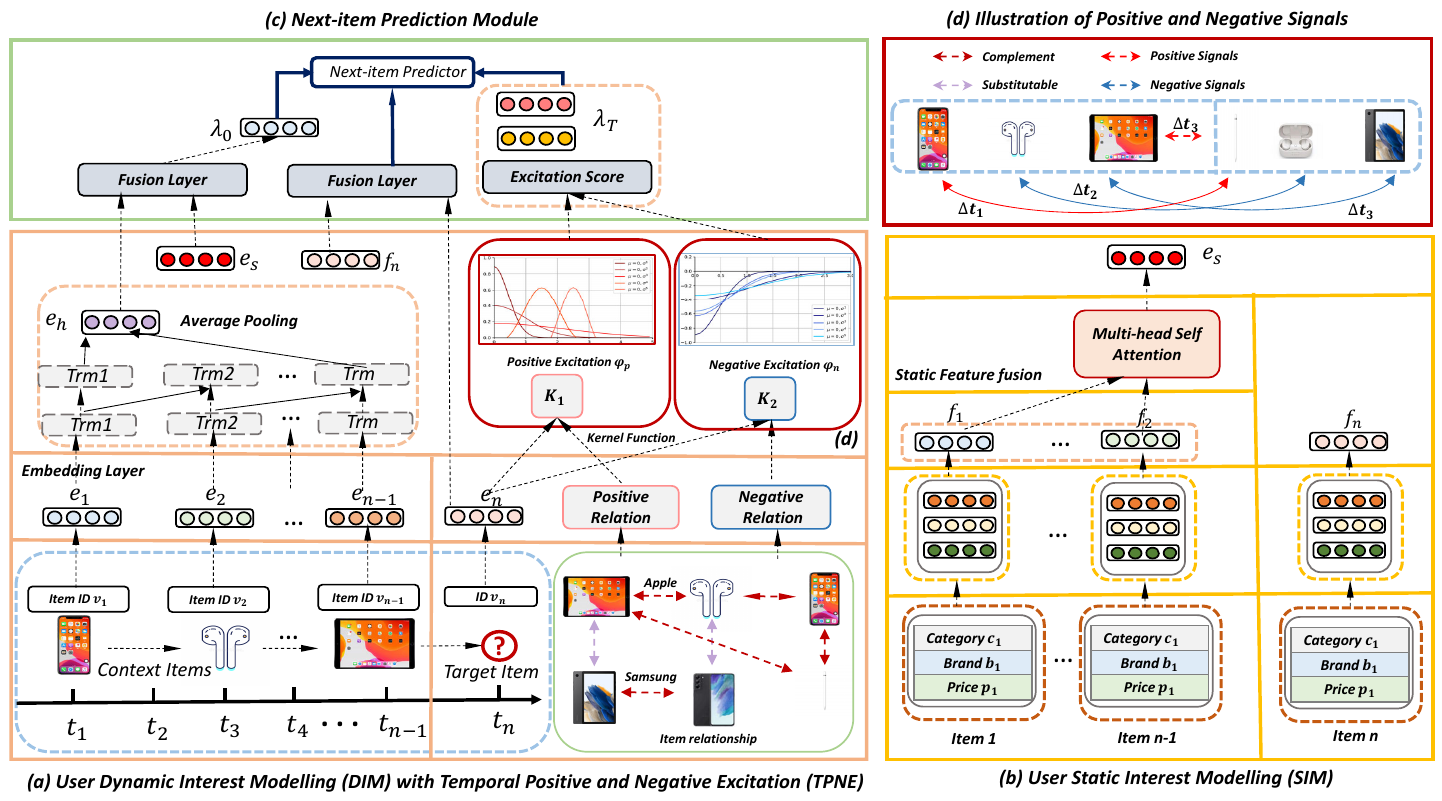}
\vspace{-2em}	
 \caption{The framework of our proposed SDIL framework. SDIL is composed of three main modules: (a) user Dynamic Interest Modeling module (DIM), (b) user Static Interest Modeling module (SIM), and (c) Next-item Prediction module. DIM captures a user's evolutionary dynamic interest by carefully modeling the time-sensitive positive and negative excitation of the user's historical interactions on the user's current interest. The SIM captures the user's relatively stable and high-level interest (e.g., interest in item category, brand) from the attribute information of interacted items. Finally, the prediction module well integrates both dynamic and static interest to obtain more precise interest for next-item prediction.}
	\label{fig:framework}
  \Description{Figure 2. Fully described in the text.}
\end{figure*}
\section{Preliminary}

\subsection{Problem Formulation}

Let $\mathcal{U} = \{u_1, u_2, ..., u_{|U|}\}$ and $\mathcal{V} = \{v_1, v_2, ..., v_{|V|}\}$ be user set and item set respectively. For each user $u \in \mathcal{U}$, there is a chronological sequence of items interacted by $u$, denoted as $H_u=\{v_1,v_2,...,v_n\}$, where $n$ is the length of the sequence $H_u$. For each item $v_i \in H_u$, it is associated with a set of attribute information including item category $c_i$, item brand $b_i$, item price $p_i$ and interaction timestamp $t_i$. Generally, for a user $u$, given her/his sequence context $O=\{v_1,v_2,...,v_{n-1}\}$ together with the associated item attribute information, the task of an SRS is to predict the next item (target item) $v_n$ which may interest $u$. 

\subsection{Hawkes Processes in Sequential Modeling}

Hawkes Processes is one of the most classic evolutionary processes \cite{hawkes,hawkes2}. The traditional Hawkes Processes is defined by the following conditional intensity function:
\begin{equation}\label{Hak}
    \lambda(t) = \mu(t) + \sum_{t_i<t}\varphi(t-t{_i}),
\end{equation}
where $\mu(t)$ represents the base intensity of the model. $t$ and $t_i$ denote the happening time of the target event and that of the last event respectively. Hence, $t-t_i$ denotes the time interval between the target event and the last event. $\varphi(\cdot)$ denotes the exciting function of a Hawkes Process. 

In practice, in the sequential recommendation scenario, a user's interest changes over time, and thus the interest evolution process can be formalized as a Hawkes Process. Accordingly, the first term $\mu(t)$ in Eq (\ref{Hak}) can be used to indicate the basic user interest in a target item $v_n$. The second term in Eq (\ref{Hak}) means the accumulative impact of all the historical context items bought before time $t$ on the user's interest at time $t$. 
\section{Methodology}

\subsection{Overview}

The framework of our proposed model is shown in Figure \ref{fig:framework}. Our model is mainly composed of (1) a temporal dynamic user interest modeling module, (2) a static user interest modeling module, and (3) a next-item prediction module. 
The temporal dynamic user interest modeling module models users' dynamic interest which changes over time by taking the ID information of historically interacted items as the input. Specifically, both the positive and negative excitation of each historical interaction on the user's next choice will be carefully considered. In addition, the temporal information (i.e., the time interval between a historical interaction and the next interaction) is well utilized to measure how positive or negative the excitation is. The static user interest modeling module models users' static interest in items based on static attribute information of items (e.g., item category, item brand) which are interacted by each user. Finally, both the learned dynamic interest and static interest are well integrated into the prediction module for the recommendation of the next item.   








\subsection{Users' Dynamic Interest Modeling}

Temporal point processes (TPPs) are a flexible and powerful paradigm that is good at modeling discrete event sequences localized in a continuous time period. A sequence of items interacted by a given user in a period well falls into this scenario. Hence, it is natural to take a TPP as the basic structure to model user-item interaction sequences. However, a classical TPP usually considers the positive excitation of each prior event (e.g., user-item interaction) on the following event (e.g., a user's next choice) and cannot model the possible negative excitation. Hence, they cannot be directly utilized to model both positive and/or negative excitation of a user's historical interactions on his/her next choice. To this end, we devise a novel Temporal Positive and Negative Excitation learning (TPNE) module to model both positive and negative excitation of historical interactions to well capture the user's dynamic interest. More importantly, TPNE is able to specify the excitation strength according to the happening time of each historical interaction.

To be specific, we take the TPP approach as a base architecture to build TPNE for capturing users' dynamic interest. The most classic model for the TPP is the Hawkes process, which is formulated as:
\begin{equation}
    \lambda_T(t) = \lambda_0 + \sum_{i:t_{i}<t_{n}}{\varphi(t_n-t_{i})},
\end{equation}
where $\lambda_0 $ denotes the user's basic interest in a candidate item. $\sum_{i:t_{i}<t_{n}}{\varphi(t_n-t_{i})}$ denotes the historical user-item interaction's excitation on target item $v_n$. For the first term, it can be calculated as:
\begin{equation}
\label{lambda}
    \lambda_0  = e_h^Te_{v_n} + u_b +i_b,
\end{equation}
where $e_h$ denotes the user historical interest embedding revealed from its historically interacted items and $e_{v_n}$ denotes the target item ID embeddings. $u_b$ and $i_b$ denote the user bias and item bias respectively. To well model the relationship between $e_h$ and $e_{v_n}$, we adopt the self-attention module to train the basic interest value $\lambda_0$. 

Given a user $u$, the input sequence is made up of item IDs sequence $H_u=\{v_1,v_2,...,v_n\}$. To obtain a unique dense embedding for each item ID, we use a linear embedding layer. Then we can get the input item embedding matrix $E \in \mathrm{R}^{|V|xd}$($|V|$ denotes the number of items and $d$ denotes the dimension of item embeddings). 
For each item, its embedding $e_{v}=E(v) \in \mathrm{R}^{1xd}$. Then we can input $H_u$'s item IDs to $E$ and obtain the user $u$'s item embedding sequences and represent it as a matrix $E(H_u)$.
Then, following previous works~\cite{AHMP,CL4Rec,SARS}, we introduce the position embedding $pos_i$ in each timestamp so as to add the position information of historical items. The input are item ID embeddings' matrix $E(H_u)$ and corresponding position embeddings $POS_i$:
\begin{equation}
    A_i = Att((E(H_u)+POS_i)W^Q, (E(H_u)+POS_i)W^K, (E(H_u)+POS_i)W^V),
\end{equation}
\begin{equation}
    Att_i = {Attention}(Q, K, V)=\operatorname{softmax}\left(\frac{Q K^{\top}}{\sqrt{d}}\right)V,
\end{equation}
where Q, K, and V denote query, key, and value respectively. $W^Q$,$W^K$ and $W^V$ denote the linear transformation matrices. 
\begin{equation}
    E(H_u) = LayerNorm(H_u + Dropout(FFN(A_i)). 
\end{equation}
Similar to previous works~\cite{AHMP,CL4Rec}, we introduce the LayerNorm, Dropout, and FFN layers to alleviate the over-fitting problem. Finally, we adopt the average pooling operation to get the user historical interest embedding $e_h$:
\begin{equation}
     e_h = \frac{1}{|N|-1}\sum_{i=1}^{n-1}{E(H_u)}_i,
\end{equation}
where $|N|$ denotes the length of the user $u$ sequence. Then we can calculate the the first term $\lambda_T$ ($e_{v_n}=E(v_n)$) as Eq. (\ref{lambda}). 

For the second term, it denotes the multi-excitation process, which represents the user history items' impact on the target item, that is to say, the dynamic user interest in target item $v_n$. Thus, it is important to curve the dynamic temporal kernel function.

Current Hawkes Process or Poisson process modeling states that all past events should have positive influences on the occurrence of current events. However, in real-world applications, there are many situations that past events give negative effects on current events. For example, after a user bought a new iPhone, in the short-term, this purchase action would give a negative excitation signal to other brand mobile phones.   
Inspired by previous work\cite{SLRS}\cite{chorus}\cite{DBLP:conf/aaai/ErtekinRM13}, we propose modeling both positive and negative excitation for SRS task, which is composed of two parts \textbf{Positive Excitation Learning} and \textbf{Negative Excitation Learning}. where
$\varphi_p$ denotes the positive factor, and $\varphi_n$ denotes the negative factor. We leverage both positive and negative signals together, making up the reactive point processes: 
\begin{equation}
\lambda_T(t)=\lambda_0 + \sum_{i: t_{i}<t_n} \varphi_p(t_n-t_{i})-\sum_{j: t_{j}<t_n} \varphi_n(t_n-t_{j}), 
\end{equation}
where $\lambda_0$ denotes the basic preference of the user on item $v_n$ as previously mentioned, $t_i$ denotes the happening time of positive interaction related to $v_n$. In contrast, $t_j$ denotes the happening time of negative interaction related to $v_n$. Then we will discuss them separately.

\subsubsection{\textbf{Positive Excitation Learning}} 

For seeking the positive correlation between historical items $O$ and target item $v_n$, we introduce four different explicit relations: also\_buy ($r_1$), also\_view ($r_2$), share\_brand ($r_3$) and similar\_item ($r_4$) contained in the datasets. For also\_buy and share\_brand relations, we regard them as complementary relations. While, for also\_view and similar\_item(two items have a similar price within the same category), we regard them as substitute relations. Then positive excitation learning can be defined as:
\begin{equation}
    \varphi_p(t_n-t_i) = \sum_{i: t_{i}<t} I_{rp}(v_i,v_n)\mathcal{K}_1(t_n-t_i),
\end{equation}
where $I_{rp}$ denotes the indicator function, if historical item $v_i$ has a relation $r \in \{r1,r2,r3,r4\}$ with $v_n$, then the $I_{rp}$=1, otherwise 0. It bridges the excitation between context information and target user interest. $\mathcal{K}_1(\cdot)$ denotes the positive temporal kernel function, which is composed of two parts as follow:
\begin{equation}
    \mathcal{K}_1^{i}(\Delta t_1)= N\left(\Delta t_1 \mid 0, \sigma_{1}^{v}\right) + N\left(\Delta t_1 \mid \mu_{2}^{v}, \sigma_{2}^{v}\right). 
\end{equation}

Previous works ~\cite{SLRS,AHMP} commonly model the short-term positive effect of historical items on target item, which only leverages the complementary information($r1$ and $r3$) to obtain the positive excitation. We also model this positive complementary relation signal in the first term  $N\left(\Delta t_1 \mid 0, \sigma_{1}^{v}\right)$, ($\Delta t_1=t_n-t_i$ represents the time interval between the target item $v_n$ and the related historical item $v_i$). $\sigma_{1}^{v}$ is decided by item ID $v_i$, which is item-specific parameter. We leverage a normal distribution function to simulate the user dynamic interest changes. For example, if a user buys a mobile phone, who may have a high possibility to buy accessories for the mobile phone in a short-term window, such as mobile phone film, matching earphones and so on. 
However, with the growth of time and the aging of the mobile phone, the interest of users to buy the corresponding supporting products would gradually decrease. Thus, we choose a mean equals to 0 normal distribution to model the user interest decay process.
For the second term, which happens because 
many items purchased by users have corresponding lifespans. When the product life cycle is just approaching, the probability of users purchasing similar substitute products will be a strong positive incentive. For example, if a user buys a mobile phone, and he/she needs to replace a brand new mobile phone one year later, then this will bring a positive excitation for the substitute relationship of similar items. We model this temporal relationship using a normal distribution with mean $\mu_{2}^{v}$, which is also related to $v_i$, which assumes the user begins to buy the substitute products after $\mu_{2}^{v}$ time interval. 

\subsubsection{\textbf{Negative Excitation Learning}} 
On the other hand, for seeking the negative correlation between historical items $O$ and target item $v_n$, we leverage two different explicit relations: also\_view ($r_2$) and similar\_item ($r_4$) contained in the datasets. We treat them as substitute relations and define the negative excitation learning as:
\begin{equation}
    \varphi_n(t_n-t_j) = \sum_{j: t_{j}<t_n} I_{rn}(v_j,v_n)\mathcal{K}_2(t_n-t_j),
\end{equation}
where $I_{rn}$ denotes the indicator function, if historical item $v_j$ has a relation $r \in \{r2, r4\}$ with $v_n$, then the $I_{rn}$=1, otherwise 0. $\mathcal{K}_2(\cdot)$ denotes the negative temporal kernel function:
\begin{equation}
    \mathcal{K}_2(\Delta t_2)=-N\left(\Delta t_2 \mid 0, \sigma_{3}^{v}\right),
\end{equation}
where $N\left(\Delta t_2 \mid 0, \sigma_{s 3}^{v}\right)$ is a normal distribution of $\Delta t$($\Delta t_2=t_n-t_j$) with 0 mean and $\sigma_3^v$ standard deviation. $\sigma_{s 3}^{v}$ is also related to $v_j$.
From an empirical analysis, users rarely buy a large number of substitutes of the same category in a short period of time. For example, a simple example, a user has just purchased a Mac laptop, so the probability of him/her buying another Lenovo laptop in the short term would be very low. However, with time goes by, the negative effect would decrease gradually, so we use a negative normal distribution with $\mu=0$ to fit this process. 
(Noting that $\mu_2^v$, $\sigma_1^v$, $\sigma_2^v$ and $\sigma_3^v$ are all learn-able parameters.)

In summary, two different temporal patterns of SRS are modeled positive excitation learning processes and negative excitation learning processes. Since these two important temporal patterns occur frequently across the whole user-item interaction history, they can provide extra-temporal information to help us capture the dynamics of users' interest.

\subsection{Users' Static Interest Modeling}

In real-world scenarios, users also have relatively fixed preferences in sequential decision-making, such as high loyalty to a certain brand and the choice of the item price, which are relatively stable in the long term and not easy to change. Therefore, we believe that this part of the more stable user interest should also be included in the modeling to achieve more accurate recommendations.

To represent the users' relatively stable interest in their interaction history, we adopt the feature-based self-attention module to model the user's static interest. Because item IDs do not clearly indicate fine-grained attributes that users are interested in, we leverage the feature-based self-attention block to search for users' preferences on static feature-level patterns. Specifically, we project the discrete and heterogeneous attributes of items into low-dimensional dense vectors and fuse these vectors, then use multi-head self-attention to model the user's static interest. Without loss of generality, we choose the category, brand, and price of three different side information to comprehensively represent different aspects of one item. We also use a linear embedding layer to obtain category embedding matrix $C \in R^{|C| \times d}$, brand embedding matrix and $B \in R^{|B| \times d}$ price embedding matrix $P \in R^{|P| \times d}$( $|C|$,$|B|$,$|P|$ denotes the number of categories, brands and price bins respectively). Furthermore, the $c_i \in R^{1 x d}$, $b_i \in R^{1 x d}$ and $p_i \in R^{1 x d}$ to denote each item different feature embeddings respectively. Our approach is easily extended to include more features. We adopt a simple yet effective additive feature fusion mode:
\begin{equation}
    f_i = c_i + b_i + p_i,
\end{equation}
where $f_i \in F$ denotes the fused feature-level item $v_i$ representation, $F$ denotes the stacked fusion matrices. Then for each user $u$, we can get his/her fused feature sequence \{$f_1$,$f_2$,$f_3$,...,$f_n$\}. We utilize the multi-head self-attention network\cite{Transformer}  to model the user's historic static interest. In the feature-level self-attention module, the query($Q$), key($K$), and value($V$) are the same and equal to $F$, and we use three different projection transformation matrices to convert them into the same subspace:  
\begin{equation}
    H_f = Att(FW^Q, FW^K, FW^V).
\end{equation}
Noting that we do not use the position here because we assume that user static interest is relatively stable and independent of user-item interaction timestamp. Where $W^Q$,$W^K$, and $W^V$ are the linear transformation matrices. Then we adopt the multi-head attention mechanism to gain more semantic information from different sub-spaces. 
\begin{equation}
    M_f = Multihead(F) = Concat(h_1, h_2, ..., h_{l_f})W^O,
\end{equation}
\begin{equation}
    h_i = Att(FW^Q_i, FW^K_i, FW^V_i),
\end{equation}
where $l_f$ denotes the length of different heads, and here we also apply the layer normalization, FFN layers and residual connection module following previous works \cite{FDSA}\cite{CL4Rec}, alleviating the over-fitting problem in the training phase.
\begin{equation}
    M_f = LayerNorm(M_f + H_f),
\end{equation}
\begin{equation}
FFN(M_f)=\operatorname{ReLU}(M_fW_{1}+b_{1})W_{2}+b_{2},
\end{equation}
where $W_1$,$W_2$ and $b_1$, $b_2$ denote the learnable parameters, we sum pooling all the $h_i$ in each user sequence to obtain the user static interest representation $e_s$.
\begin{equation}
    e_s = \frac{1}{|N|-1}\sum_{i=1}^{|N|-1}{H_f},
\end{equation}
$e_s$ provides more stable user interest on certain attributes (i.e. brand, price in our case), such characteristics may not be easily perceived but are indeed potential consumption habits of users.

\subsection{Next-item Prediction}

Since both dynamic user interest and static user interest are vital in SRS task, it is crucial to combine them and as the input of the downstream prediction module for
next-item prediction. Hence, inspired by previous works~\cite{MaMZSLC20, DSIM}, we leverage the gate fusion module to fuse the dynamic embedding $e_h$ and static interest embedding $e_s$ so as to better next-item prediction. 
\begin{equation}
    g = \sigma(W_1e_s+W_2e_h + b),
\end{equation}
\begin{equation}
\label{e_fusion}
    e_{f} = g \odot e_s + (1-g) \odot e_h,
\end{equation}
where $W_1$,$W_2$, and $b$ are the learnable parameters in the gating layer. $\odot$ denotes the element-wise multiplication and $g \in \mathcal{R}^{1xd}$ denotes the learnable gate. Since there is some redundant information between user dynamic interest and static interest, we adaptively pass the informative messages and restrain the useless ones to obtain the final interest representation $e_{f}$.
In order to learn the parameters of the SDIL framework, we adopt the pairwise ranking loss (BPR loss)~\cite{BPR} to optimize our model: 
\begin{equation}
    \mathcal{L}_{r}=-\sum_{u \in \mathcal{U}} \sum_{i=1}^{N_{u}} \log \sigma\left(\hat{y}_{u i}-\hat{y}_{u j}\right),
\end{equation}
\begin{equation}
   \hat{y}_{u i} = e_{f}^Te_i + \lambda_{T,i},
\hspace{2ex}
    \hat{y}_{u j} = e_{f}^Te_j+ \lambda_{T,j}
\end{equation}
where $\sigma$ represents the sigmoid function, $\hat{y}_{u i}$ represents the final preference score of user $u$ to positive item $i$ while $\hat{y}_{u j}$ represents the final preference score of user $u$ to negative item $j$.

\section{Experiments}

\subsection{Experimental Setting}

\subsubsection{ \textbf{Dataset Preparation and Baselines}}
We conduct experiments on publicly accessible Amazon datasets \cite{AmazonDataset}.
We choose three representative sub-datasets from Amazon datasets: \textit{Beauty}, \textit{Cell Phones and Accessories} (Cellphones) and \textit{Toys and Games} (Toys) and keep the ‘5-core’ datasets \cite{S3Rec}\cite{FPMC}\cite{ICL}, which filter out user-item interaction sequences with length less than 5.
The statistics of evaluation datasets are shown at Table \ref{tab:datasets} in Appendix A. Following the previous works\cite{chorus}\cite{SLRS}, we take "also buy" as complementary relations and "also view" as substitute relations between items, i.e., items $v_1$ and $v_2$ are complementary if most users who buy $v_1$ also buy $v_2$. In addition, we introduce two more effective item relationships: (1) items of the same brand have complementary relations (e.g., iPhone and AirPod), and (2) items of the same category with similar prices have substitute relation (e.g., Canon cameras and Nikon cameras). Furthermore, we also take some item attribute information into account, including fine-grained item category, item brand and item price. Item category and brand are categorical features and we use the unique one-hot encoding to represent them. For price information, we bin and cut them into 10 intervals.    

For reproducibility, we follow the commonly used benchmark setting of ReChorus ~\cite{chorus,SLRS} to set up our experiments. Specifically, for each user, we first discard duplicated interaction sequences and sort the items in each user's sequence chronologically by their timestamp. 
Furthermore, the maximum length of interaction sequences is set to 20. If there are more than 20 interactions in a sequence, we adopt the latest 20 interactions. If the number of interactions in a sequence is less than 20, we make it up to 20 by padding virtual items with the ID of 0. For all baselines on all three experimental datasets, hidden size and batch size is set to 64. Following the common practice in sequential recommendations, we leave the interactions happening at the latest time as the test dataset and the interactions at the second latest time as the validation dataset.
To evaluate the performance of our model TPNE, we select 10 representative and/or state-of-the-art recommendation models as baselines. The detailed information are listed in the Appendix \ref{baselines}.

\begin{table*}[ht]
    \small
     \caption{Overall performance. Bold scores represent the highest results of all methods. Underlined scores stand for the second-highest results. Our model achieves the state-of-the-art result among all baseline models. $^*$ means the improvement is significant at $p < 0.05$.}
     \begin{tabular}{c|l|c|cccc|cccc|c|c|c}
     \toprule
          Dataset& Metric      &BPR & GRU4Rec & Caser &NARM & SASRec & TiSASRec & SLRS+ &Chorus &AHMP &KDA & SDIL & Improv.\\
          \midrule                  
          \multirow{9}*{Beauty}&HR@5&0.3317&0.3202&0.3210&0.3334&0.3666&0.3872&0.4339&0.4536&0.4566&\underline{0.4860}&$\textbf{0.4926}^*$&1.36\%\\
                               &HR@10&0.4355&0.4311&0.4345&0.4462&0.4590&0.4559&0.5337&0.5698&0.5519&\underline{0.5997}&$\textbf{0.6128}^*$&2.18\%\\
                               &HR@20&0.5505&0.5693&0.5757&0.5823&0.5743&0.5700&0.6361&0.6838&0.6599&\underline{0.7144}&$\textbf{0.7323}^*$&2.51\%\\
    
                               &NDCG@5&0.2361&0.2271&0.2246&0.2348&0.2797&0.2904&0.3319&0.3386&0.3496&\underline{0.3648}&$\textbf{0.3698}^*$&1.37\%\\
                               &NDCG@10&0.2697&0.2628&0.2612&0.2712&0.3094&0.3036&0.3642&0.3762&0.3803&\underline{0.4016}&$\textbf{0.4088}^*$&1.79\%\\
                               &NDCG@20&0.2987&0.2976&0.2967&0.3055&0.3385&0.3324&0.3900&0.4050&0.4076&\underline{0.4306}&$\textbf{0.4390}^*$&1.95\%\\
    
                               &MRR&0.2363&0.2271&0.2246&0.2366&0.2923&0.2904&0.3319&0.3386&0.3421&\underline{0.3549}&$\textbf{0.3610}^*$&1.72\%\\

         \midrule          
         \multirow{9}*{Cellphone}&HR@5&0.3387&0.3015&0.3937&0.4168&0.4439&0.4520& 0.4696&0.4697&0.5045&\underline{0.5497}&$\textbf{0.5538}^*$&0.75\%\\
                               &HR@10&0.4528&0.4301&0.5309&0.5509&0.5595&0.5767& 0.5641&0.5929&0.6132&\underline{0.6745}&$\textbf{0.6792}^*$&0.70\%\\
                               &HR@20&0.5852&0.5918&0.6810&0.6974&0.6817&0.7022& 0.6637&0.7152&0.7284&\underline{0.7923}&$\textbf{0.8028}^*$&1.33\%\\
                               
                               &NDCG@5&0.2430&0.2085&0.2800&0.2995&0.3353&0.3344& 0.3634&0.3530&0.3852&\underline{0.4119}&$\textbf{0.4188}^*$&1.69\%\\
                               &NDCG@10&0.2798&0.2498&0.3243&0.3429&0.3727&0.3748&0.3939&0.3929&0.4204&\underline{0.4523}&$\textbf{0.4595}^*$&1.59\%\\
                               &NDCG@20&0.3131&0.2905&0.3622&0.3799&0.4036&0.4065& 0.4191&0.4238&0.4495&\underline{0.4821}&$\textbf{0.4908}^*$&1.80\%\\
    
                               &MRR    &0.2453&0.2271&0.2246&0.2969&0.2923&0.2904&0.3319&0.3386&0.3747&\underline{0.3666}&$\textbf{0.4049}^*$&10.45\%\\

         \midrule       
         \multirow{9}*{Toys}   &HR@5&0.2897&0.2902&0.2898&0.3173&0.3602&0.3475& 0.4368&0.4124&0.4603&\underline{0.4805}&$\textbf{0.4953}^*$&3.08\%\\
                               &HR@10&0.3897&0.4060&0.4103&0.4336&0.4570&0.4608& 0.5345&0.5203&0.5587&\underline{0.5882}&$\textbf{0.6069}^*$&3.18\%\\
                               &HR@20&0.5061&0.5546&0.5590&0.5777&0.5700&0.6003& 0.6440&0.6443&0.6621&\underline{0.7019}&$\textbf{0.7248}^*$&3.26\%\\
                               
                               &NDCG@5&0.2068&0.1974&0.1947&0.2206&0.2738&0.2535& 0.3490&0.3132&0.3600&\underline{0.3660}&$\textbf{0.3797}^*$&3.74\%\\
                               &NDCG@10&0.2390&0.2348&0.2336&0.2581&0.3050&0.2901&0.3804&0.3480&0.3918&\underline{0.4007}&$\textbf{0.4157}^*$&3.74\%\\
                               &NDCG@20&0.2683&0.2721&0.2710&0.2944&0.3334&0.3253& 0.4081&0.3793&0.4179&\underline{0.4294}&$\textbf{0.4454}^*$&3.73\%\\
                               &MRR    &0.2116&0.2271&0.2246&0.2244&0.2923&0.2904&0.3319&0.3386&0.3547&\underline{0.3666}&$\textbf{0.3713}^*$&1.28\%\\
          \bottomrule
     \end{tabular}
     \label{tab:overall}
    \end{table*}

\subsubsection{ \textbf{Implementation Details and Parameter Settings}}
To ensure fair comparisons, we follow the ReChorus' baseline implementation\footnote{https://github.com/THUwangcy/ReChorus}. Specifically, for both our model and all baseline models, we set the embedding size and batch size to 64. The number of training epochs for all models is set to 150 and an early-stop strategy is used. When a model's performance on the validation set decreases for 10 consecutive rounds, the training will stop. Other parameter settings for each baseline are specified in Appendix \ref{Baseline Parameter Settings}. 

For our SDIL model, we set the number of heads in the multi-head attention model of SIM to 4 and the number of layers of both DIM and SIM to 2 by using a grid search from 1 to 8 with a step of 1. We set the embedding dimension of all the attributes including price, brand, and category to 64. Our model is implemented by \textit{PyTorch}. we pre-trained the item embedding with a learning rate of 5e-4 and the main model is optimized by Adam optimizer with a learning rate of 1e-4. 

\begin{figure*}[htp]
\centering

\begin{subfigure}[t]{.33\linewidth}
  \centering
\includegraphics[width=\linewidth,scale=1.00]{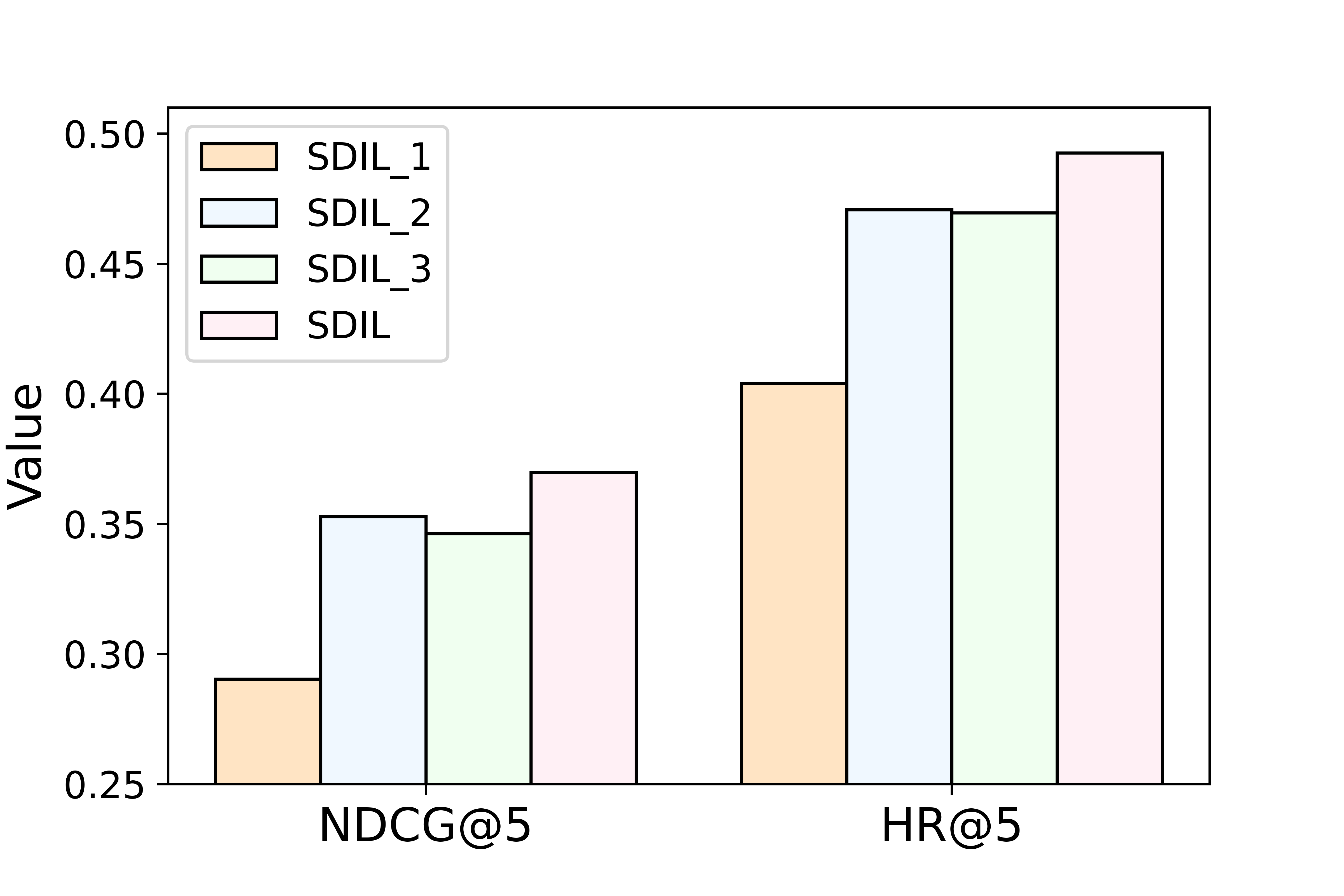}
	\caption{Beauty Dataset}
	\label{fig:Beauty}
\end{subfigure}
\begin{subfigure}[t]{.33\linewidth}
  \centering
 	\includegraphics[width=\linewidth,scale=1.00]{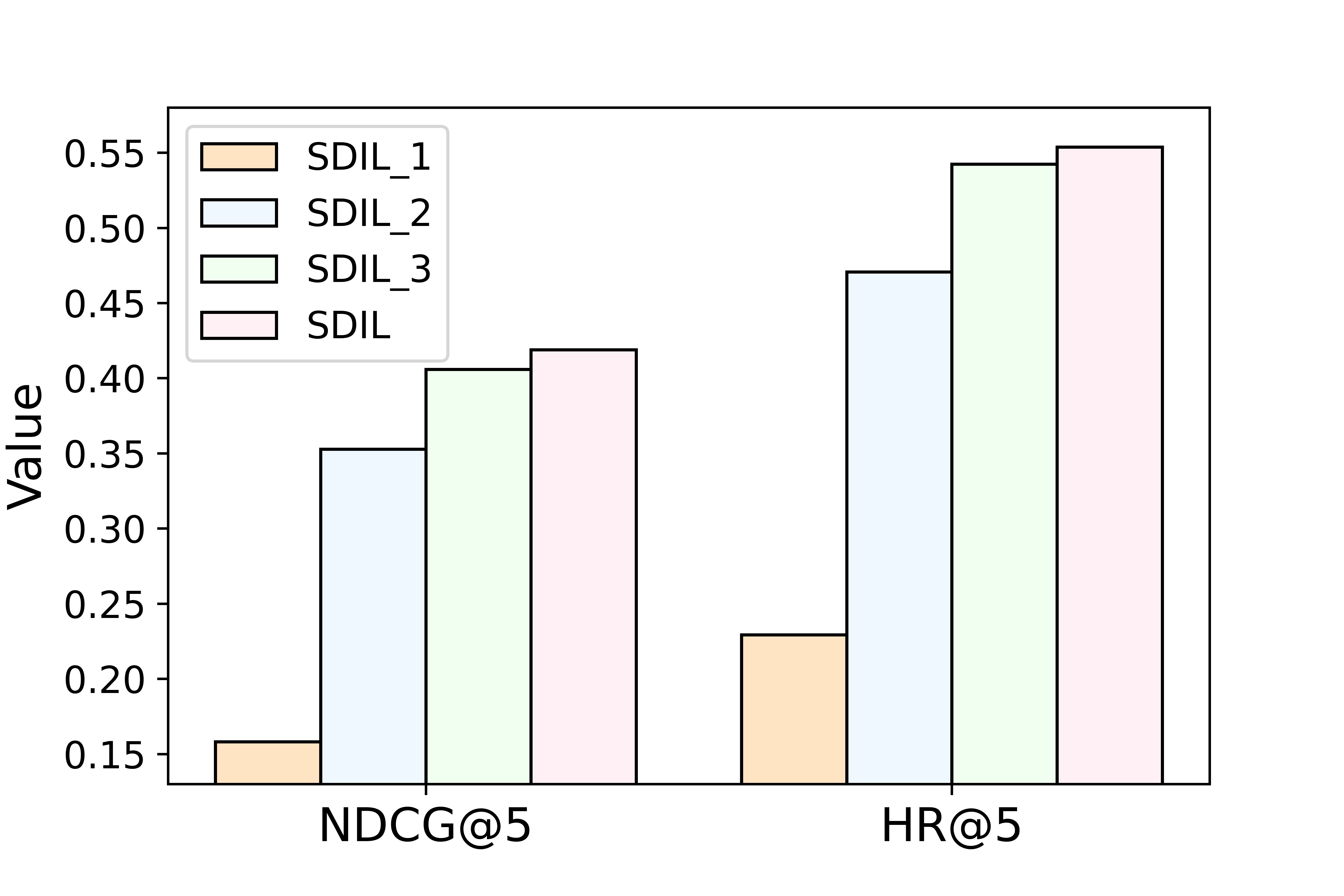}
	\caption{Cellphones Dataset}
	\label{fig:Cellphones}
\end{subfigure}
\begin{subfigure}[t]{.33\linewidth}
  \centering	\includegraphics[width=\linewidth,scale=1.00]{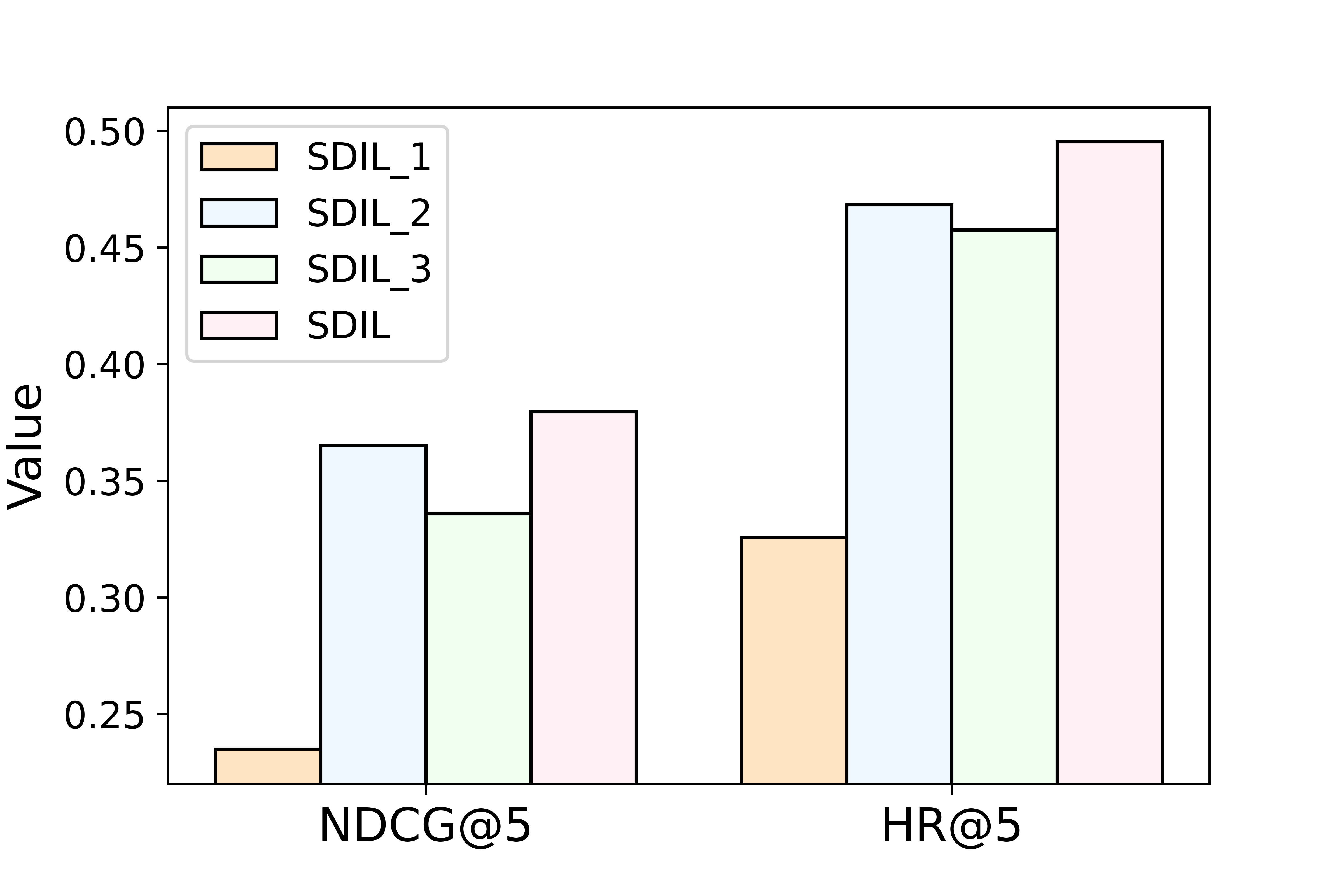}
    \caption{Toys Dataset}
	\label{fig:Toys}
\end{subfigure}
\vspace{-0.4cm}
\caption{Ablation study on the model performance (HR@5 and NDCG@5) on different datasets.}
\label{fig:ablationstudy}
  \Description[fig:Beauty, fig:Cellphones and fig:Toys look identical]{fig:Beauty, fig:Cellphones and fig:Toys comparison shows identical length, wingspan, and overall bodily structure.}

\end{figure*}

\subsubsection{ \textbf{Evaluation Metrics}}

We use three commonly used metrics including \textit{Normalized Discounted Cumulative Gain @K} (NDCG@K), \textit{Hit Ratio @K} (HR@K) and \textit{Mean Reciprocal Rank} (MRR)~\cite{SARS,chorus} to evaluate the performance of all compared methods.
For all the baseline models, we generate the ranking list of items for each testing interaction. Both of them are applied with $K$ chosen from \{5, 10, 20\}. In order to accelerate the evaluation, following the work \cite{chorus}, we evaluate the ranking results with 99 randomly selected negative items. Meanwhile, we follow the setting in\cite{WangRMCMR19}, using the paired t-test with p<0.05 for the significance test.

\subsection{Performance Comparison with Baselines}

Table \ref{tab:overall} reports the comparison results between our method and 10 different baseline methods on three datasets. 
The best results and the second-best results across all methods are in bold and underlined respectively. We have the following observations:

(1) First, we can find that all the non-time-sensitive sequential methods such as GRU4Rec~\cite{GRU4Rec1,GRU4Rec2}, Caser~\cite{Caser} surpass the non-sequential method BPR, which indicates the effectiveness of modeling sequential information. Meanwhile, NARM utilizes both global and local encoders to model the long- and short-term interest of users and thus performs better than GRU4Rec and Caser. Moreover,
the self attention-based SRS, i.e., SASRec~\cite{SARS}, consistently outperforms previous traditional deep learning methods including CNN-based Caser and RNN-based GRU4Rec, proving the effectiveness of using the attention mechanism to encode sequence data. 

(2) Second, more advanced time-sensitive sequential models, such as TiSASRec~\cite{TiSASRec}, generally inherit the attention-based encoder while introducing extra-temporal signals to further improve the recommendation performance. Specifically, TiSASRec explicitly models both the absolute positions of items as well as time intervals between them in a sequence, and thus it outperforms SARS w.r.t most evaluation metrics on most datasets. SLRS+~\cite{SLRS} introduces both knowledge-graph information and the multi-excitation of items to further enhance the recommendation performance. Chorus~\cite{chorus} further models the different situations of the temporal effect of user-item interactions on item embedding level and thus outperforms SLRS+. However, due to the limited representation ability of its utilized base model, Chorus shows the over-fitting issue. AHMP~\cite{AHMP} models the user temporal category-level consumption excitation via self-attention models, achieving the third-best results on most datasets. KDA~\cite{KDA} designs the virtual relations between items to enhance the relation learning between items, which helps KDA obtain a significant improvement over other baselines. 

(3) Third, our proposed SDIL model consistently outperforms all baselines in all datasets. The average improvement on the Cellphones dataset compared with the best-performing baseline is 10.45\% w.r.t MRR. The reason is obvious. Different from most of the existing SRS methods, which ignore modeling either users' relatively stable static interest, or the negative excitation of historical interactions on user's next choice, our proposed SDIL model has been carefully designed to well address these issues. To be specific, SDIL not only well models users' static interest in items via carefully capturing users' interest on item attribute level, but also comprehensively model users' dynamic interest via the well-designed TPNE module. More importantly, TPNE is able to model both positive and negative excitation of users' historical interactions on their next choice, leading to significant performance improvement.             

\subsection{Ablation Study}
To analyze the rationality and effectiveness of our designed components in the SDIL model, we conduct the ablation study with three variants. \textbf{SDIL-1}: Only uses static interest module to predict the next item; \textbf{SDIL-2}: Only uses dynamic interest module to predict the next item; \textbf{SDIL-3}: Uses both dynamic and static interest module to predict the next item without modeling excitation. 
Figure \ref{fig:ablationstudy} presents the performance of these three variants and the full model \textit{SDIL} on the three experimental datasets. First, compare \textit{SDIL-1} and SDIL, we find the performance of \textit{SDIL-1} significantly decreases when dropping the dynamic interest modeling module TPNE from \textit{SDIL}. This proves the extremely importance of our designed TPNE module.
Second, the comparison between \textit{SDIL-2} and \textit{SDIL} demonstrates that static interest modeling is also vital for recommendation performance. The performance clearly decreases without it. Third, the comparison between \textit{SDIL-3} and \textit{SDIL} demonstrates the importance of modeling historical interactions' excitation on users' next choice. In summary, the dropping of any of our designed components would lead to a clear reduction in recommendation performance, this effectively verifies the effectiveness and rationality of our design. 

\subsection{The Effectiveness of Modeling Negative Excitation} To verify the rationality of incorporating negative excitation and the effectiveness of our proposed TPNE module, we keep the SDIL framework while changing its TPNE module to TPE to obtain another variant SDIL-TPE. TPE models users' dynamic interest by only considering temporal \textit{positive} excitation. The result is provided in Appendix \ref{neg_effectiveness}. As Table \ref{tab:kernel functions} shows, SDIL consistently surpasses SDIL-TPE, indicating that the comprehensive modeling of both positive and negative excitation is more effective compared with the modeling of only positive excitation.

\subsection{Parameter Sensitivity Test \& Case Study}
We conduct the parameter sensitivity test to see how the key model parameters will affect the performance of our SDIL model. The result is shown in Appendix \ref{Sensitivity} to save space. In addition, we conduct case studies to show some insights on how our proposed model can work better in a straightforward way. Due to the limited space, the details are provided in Appendix \ref{case_study}.   

\section{Conclusion}

In this paper, we propose a novel Static-Dynamic Interest Learning (SDIL) framework to comprehensively model users’ static interest and dynamic interest in items for generating accurate sequential recommendations. More importantly, to comprehensively model users' dynamic interest, we devise a novel Temporal Positive and Negative Excitation modeling (TPNE) module. TPNE can not only well capture both the positive and negative excitation of each user's historical interactions on his/her next choice of items, but also is able to specify the excitation strength for each historical interaction according to its happening time. Our extensive experimental results on three real-world datasets validate the superiority of our proposed model over state-of-the-art methods. In the future, we will explore how to better model users' static interest to further enhance the performance of our model.

\bibliographystyle{ACM-Reference-Format}
\bibliography{sample-base}

\appendix
\clearpage

\section{Data Description}

\begin{table}[h]
    \caption{Statistics of the datasets after preprocessing.}
    \small
    \begin{tabular}{l|cccc}
    \toprule
    Specs. & Beauty & Cellphones & Toys   \\ 
    \midrule
    \# Users &  22,363 & 27,879 & 19,412\\
    \# Items & 12,101 & 10,429 &  11,924 \\
    \# Avg. Seq Length & 8.8 & 7.0 & 8.6 \\
    \# Interactions &  198,502 &194,439 & 167,597  \\
    \# Sparsity & 99.93\%  & 99.94\% & 99.93\%   \\
    \# Item Categories & 148 & 20  & 145   \\
    \bottomrule
    \end{tabular}
    \label{tab:datasets}
\end{table}

\section{The Effectiveness of Negative Excitation} \label{neg_effectiveness}

To verify the rationality of incorporating negative excitation and the effectiveness of our proposed TPNE module, we keep the SDIL framework while changing its TPNE module to TPE to obtain another variant SDIL-TPE. TPE models users' dynamic interest by only considering temporal \textit{positive} excitation only. As Table \ref{tab:kernel functions} shows, SDIL consistently surpasses SDIL-TPE, indicating that the comprehensive modeling of both positive and negative excitation is more effective compared with the modeling of only positive excitation. 
\begin{table}[h]
  \caption{Performance comparison between SDIL-TPE and SDIL. $^*$ means the improvement is significant at $p < 0.05$.}
  \small
  \label{tab:kernel functions}
  \setlength{\tabcolsep}{1.3mm}{
  \begin{tabular}{c|l|cc}
    \toprule
    \multirow{1}{*}{Dataset} & \multirow{1}{*}{Metrics} & 
    \multirow{1}{*}{SDIL-TPE} & 
      \multirow{1}{*}{\textbf{SDIL}} \\
    \hline
    \midrule
    \multirow{5}{*}{Beauty} 
    & \textit{HR@5}    & 0.4825 & \textbf{$0.4926^*$}  \\
    & \textit{HR@10}   & 0.6054 & \textbf{$0.6128^*$}  \\
    & \textit{NDCG@5}  & 0.3487 & \textbf{$0.3698^*$}  \\
    & \textit{NDCG@10} & 0.4014 & \textbf{$0.4088^*$}  \\
    & \textit{MRR}     & 0.3534 & \textbf{$0.3602^*$}  \\
    \midrule
    \multirow{5}{*}{Cellphones}   
     & \textit{HR@5}   & 0.5521 & \textbf{$0.5538^*$} \\
    & \textit{HR@10}   & 0.6772 & \textbf{$0.6792^*$} \\
    & \textit{NDCG@5}  & 0.4102 & \textbf{$0.4188^*$} \\
    & \textit{NDCG@10} & 0.4422 & \textbf{$0.4595^*$} \\
    & \textit{MRR}     & 0.4038 & \textbf{$0.4049^*$} \\
    \midrule
    \multirow{5}{*}{Toys}   
    & \textit{HR@5}      & 0.4871 & \textbf{$0.4953^*$}  \\
    & \textit{HR@10}     & 0.5979 & \textbf{$0.6069^*$}  \\
    & \textit{NDCG@5}    & 0.3741 & \textbf{$0.3797^*$}  \\
    & \textit{NDCG@10}   & 0.3670 & \textbf{$0.4157^*$}  \\
    & \textit{MRR}       & 0.3670 & \textbf{$0.3713^*$}  \\
  \bottomrule
\end{tabular}}
\end{table}

\section{Parameter Sensitivity Analysis \label{Sensitivity}}
In this section, we will discuss the parameter setting's effect on the model performance. The results are presented in Figure \ref{fig_comparison}. The embedding size of item representation does affect the performance of SRS models. We can see that the performance of the model increases consistently as the embedding size increases.
\begin{figure}[!h]
\centering
\begin{subfigure}[t]{.48\linewidth}
  \centering
	\includegraphics[width=\linewidth,scale=1.00]{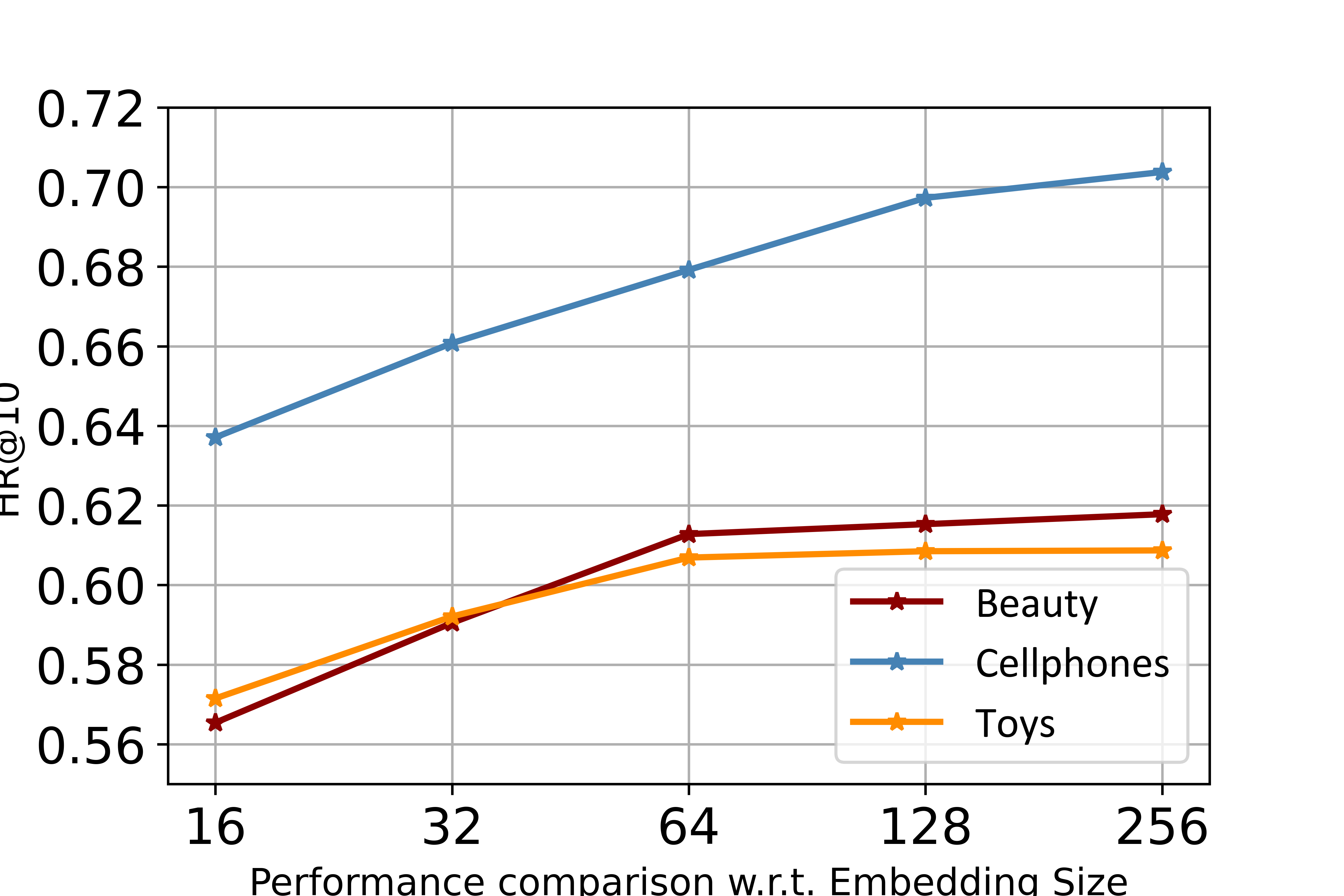}
	\label{fig:Embedding_Size_HR}
\end{subfigure}
\hfill
\begin{subfigure}[t]{.48\linewidth}
  \centering
 	\includegraphics[width=\linewidth,scale=1.00]{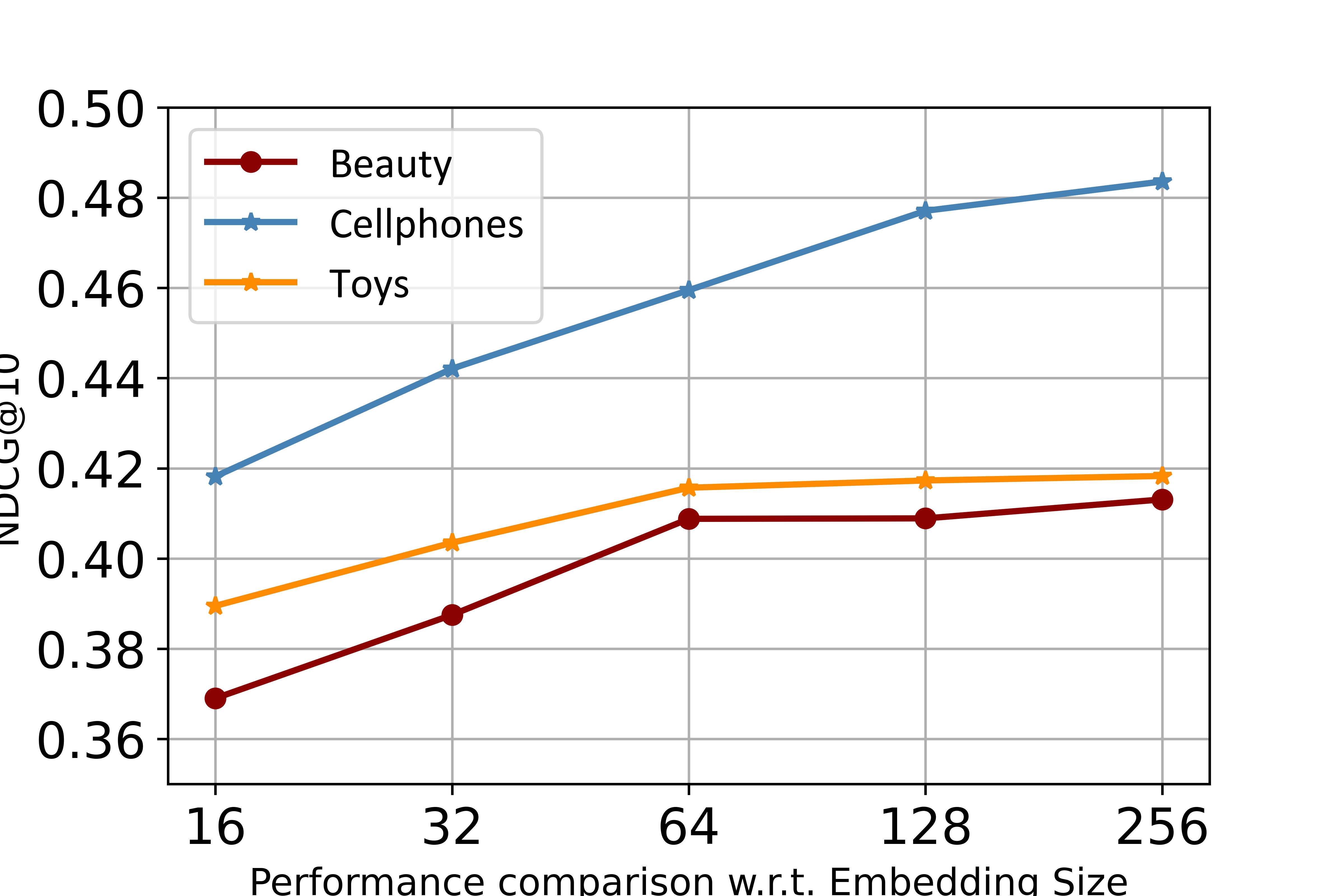}
	\label{fig:Embedding_Size_NDCG}
\end{subfigure}
\hfill
\vspace{-0.5cm}
\caption{Embedding size setting's effect on the model performance. (HR@5 and NDCG@10).}
\label{fig_comparison}
\Description[fig:Embedding Size HR and fig:Embedding Size NDCG look identical]{fig:Embedding Size HR and fig:Embedding Size NDCG comparison shows identical length, wingspan, and overall bodily structure.}
\vspace{-0.3cm}
\end{figure}

Besides, we also conduct experiments about the different multi-head numbers and transformer layers. The head number of TPNE layers is searched from \{1,2,4,6\} and the Transfomer layer number is searched from \{0,1,2,3,4\}. We can find that the performance increases as the head number increases, while when the layer or head number is too large, the model begins to over-fit.
\begin{figure}[!h]
\centering
\begin{subfigure}[t]{.48\linewidth}
  \centering
	\includegraphics[width=\linewidth,scale=1.00]{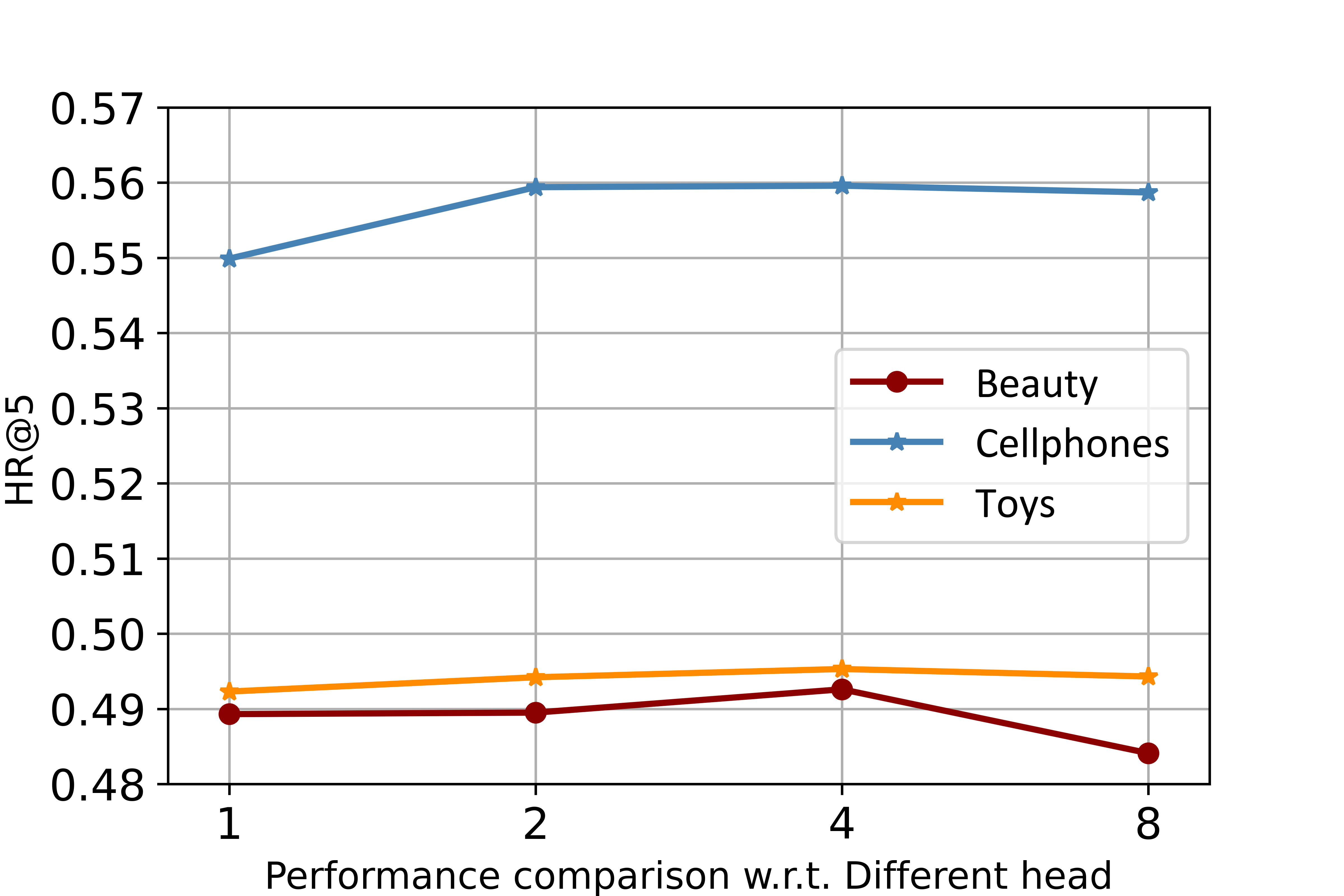}
	\label{fig:multi-head_numbers}
\end{subfigure}
\hfill
\begin{subfigure}[t]{.48\linewidth}
  \centering
 	\includegraphics[width=\linewidth,scale=1.00]{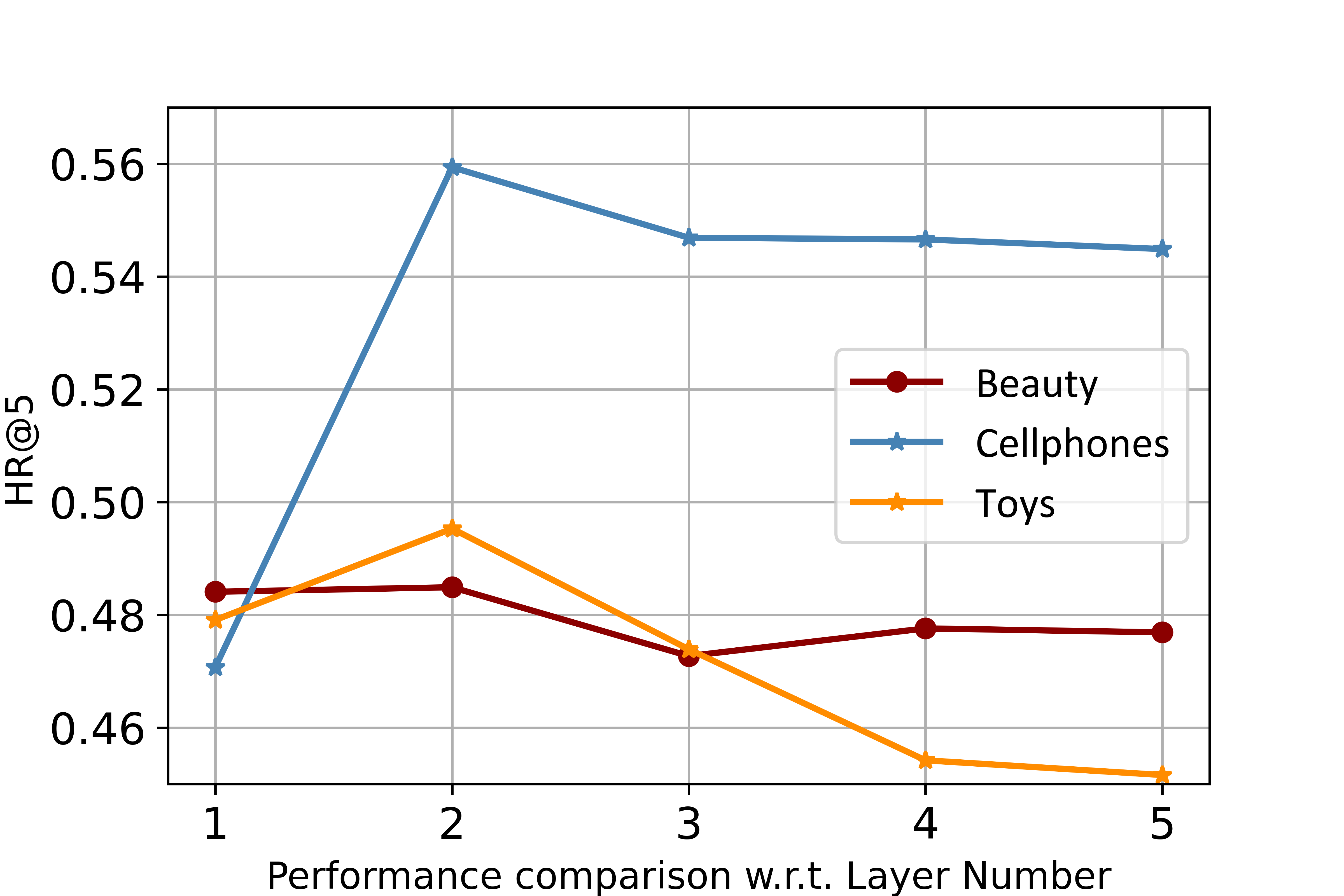}
	\label{fig:layer_numbers}
\end{subfigure}
\hfill
\caption{Different transformer layers setting's effect on the model performance. (HR@10 and NDCG@10).}
\label{fig_comparison}
\Description[fig:multi-head numbers and fig:layer numbers look identical]{fig:multi-head numbers and fig:layer numbers comparison shows identical length, wingspan, and overall bodily structure.}

\end{figure}

\section{Case Study \label{case_study}}

\begin{figure}[!h]
	\centering
	\includegraphics[scale=0.3]{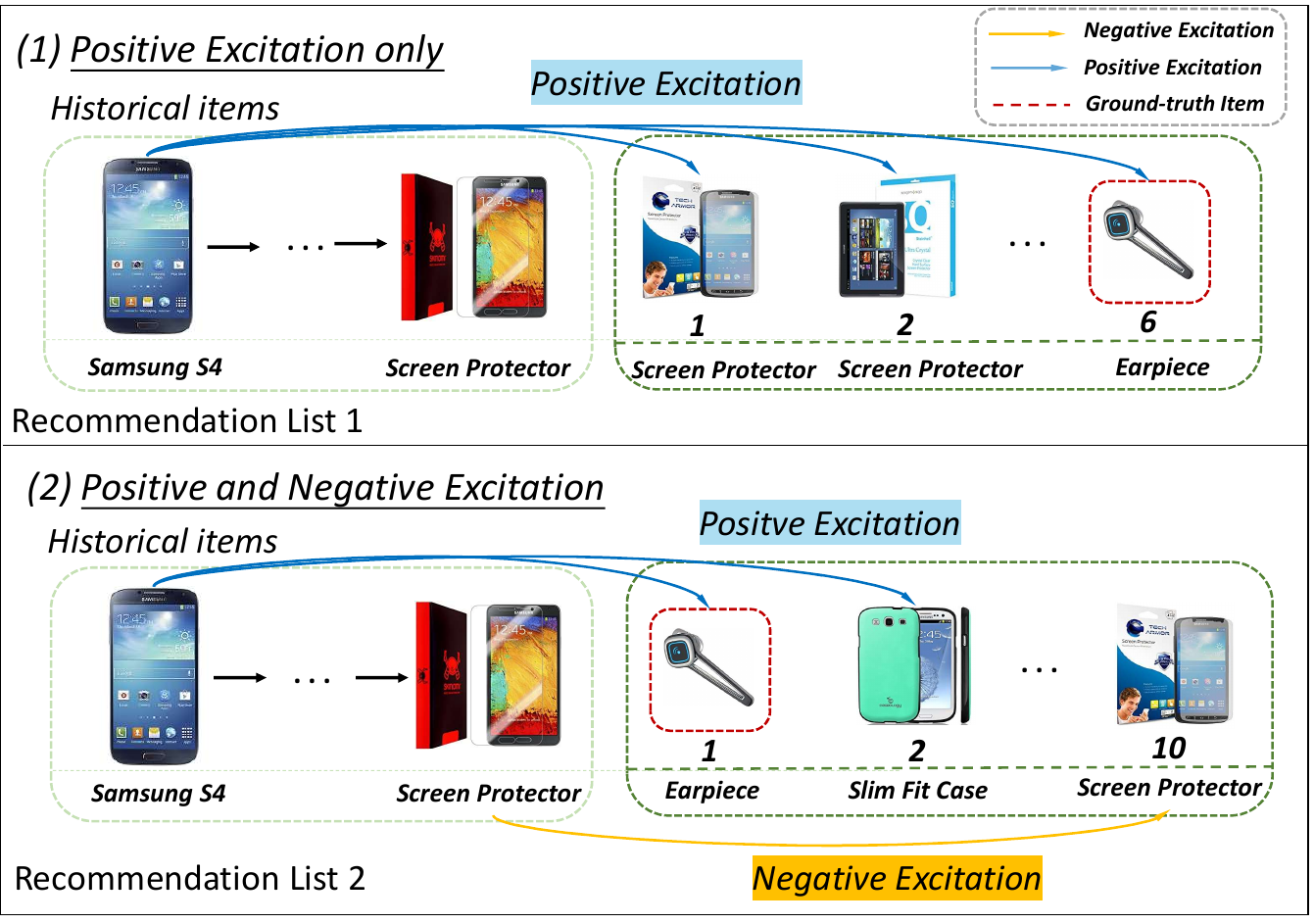}
         \vspace{-0.2cm}
    \caption{Illustration of the ranking results of TPE and TPNE. The item highlighted in the red boxes is the ground-truth item.}
	\label{fig:case study2}
    \Description{Figure 6. Fully described in the text.}

\end{figure}

\begin{figure}[!h]
	\centering
	\includegraphics[scale=0.3]{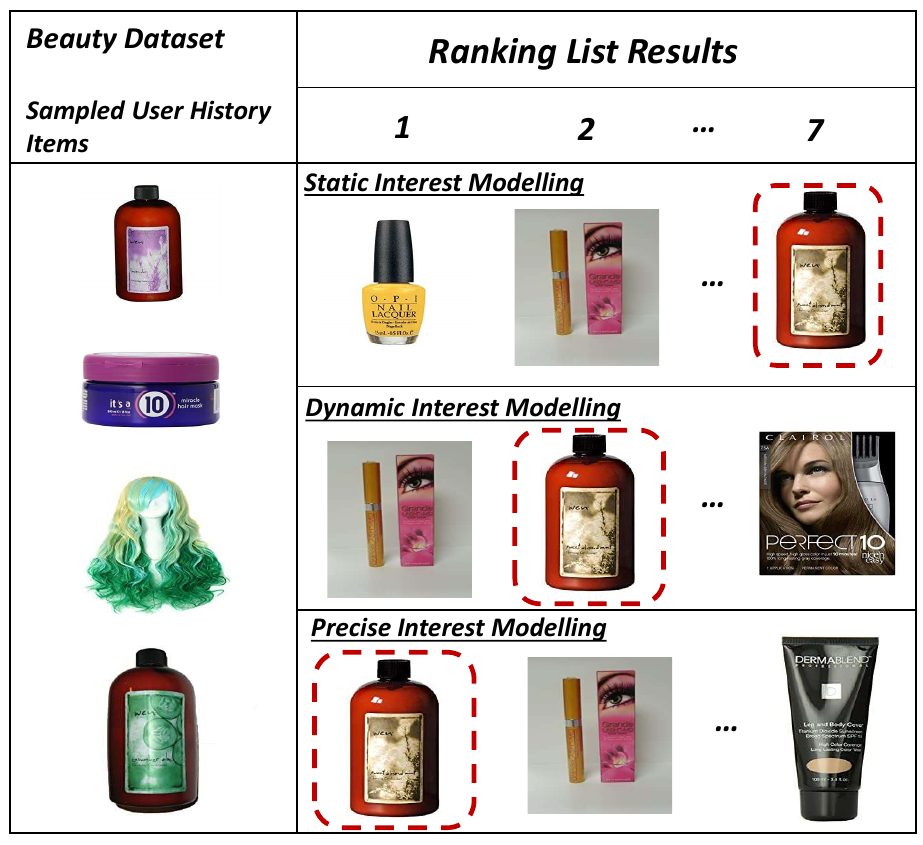}
        \vspace{-0.2cm}
	\caption{Illustration of the ranking results of SIM, DIM and DSIM. The item highlighted in the red boxes is the ground-truth item.}
	\label{fig:case study}
    \Description{Figure 7. Fully described in the text.}
\end{figure}

To get a better understanding of the effectiveness of our TPNE in a specific context, we randomly select a user sequence and list the recommendation results of TPNE and its variant TPE. In Figure \ref{fig:case study2}, the user's interaction history demonstrates that the user first purchased the Samsung S4 phone and then purchased the corresponding accessories, including the wallet case holder, car phone holder, and screen protector. For the next-item prediction task, the ground-truth item is a blue-tooth earpiece. For the TPE using only positive incentive learning, although it can capture the positive excitation between the phone and the earpiece, it also introduces the positive excitation with the screen protector. However, the user has already purchased the screen protector in the previous moment, so it will lead to homogeneous recommendations, resulting in inaccurate recommendation results. As for TPNE, our model considers the negative excitation of similar items in the short term, in this case, for the similar screen protector items in timestamp $t$, the model will come out with a negative mutual exclusion signal. Thus, the homogeneous recommendation can be effectively alleviated, and better recommendation performance can be obtained. It verifies the importance of considering both positive and negative excitation learning. Moreover, to prove the superiority of SDIL in combining both dynamic and static interest contexts. 
We listed the ranking results of DIL, SIL, and SDIL. As Figure \ref{fig:case study} illustrated, we random sample one user interaction item history. We can also find our SDIL can well capture the user intent in the next-item prediction task. 

\section{Baselines for Comparisons}
\label{baselines}
To evaluate the performance of our model TPNE, we select 10 representative and/or state-of-the-art recommendation models as baselines. They are classified into the following three groups: \textbf{(1) Non-sequential models:} \textit{BPR}~\cite{BPR}: This model characterizes the pairwise interactions via a matrix factorization model and optimizes through a pair-wise Bayesian Personalized Ranking loss.
\textbf{(2) Non-time-sensitive sequential models:} \textit{GRU4Rec}~\cite{GRU4Rec1}: This model uses the GRU to model the user interaction sequence and gives the final recommendation. \textit{NARM}~\cite{NASR}: It is a classic long- and short-term user interest modeling, which designs two RNN-based global encoders and a local encoder to model the user's long and short-term interest respectively. \textit{Caser}~\cite{Caser}: This model embeds items in user interaction history as image form, using convolutional filters for the recommendation. \textit{SASRec}~\cite{SARS}: This model leverages users’ longer-term semantics as well as their recent actions simultaneously for accurate next-item recommendations.
\textbf{(3) Time-sensitive sequential models:}
\textit{TiSASRec}~\cite{TiSASRec}: This model leverages the timestamp in the user-item interaction, exploring the different time intervals' influence in the next item prediction. \textit{SLRS+}~\cite{SLRS}: SLRS combines Hawkes process and MF into one framework, which aims at modeling the user repeat consumption in the sequential recommendation. Since the Amazon dataset removes the repeat consumption in the test set, SLRS+ uses the Hawkes process to model the relations including also view and also buy behaviors. \textit{AMHP}~\cite{AHMP}: This model introduces the repeat category consumption into sequential recommendation and adopts the self-attention architecture as the main model. \textit{Chorus}~\cite{chorus}: This model is a state-of-the-art method with item relations and temporal evolution. \textit{KDA}~\cite{KDA}: KDA devises relational intensity and frequency-domain embeddings to adaptively determine the importance of historical interactions.

\section{Baseline Parameter Settings}\label{Baseline Parameter Settings}
For fair comparison, we carefully tuned the model parameters of all the baselines on the validation set. Then we choose the model parameters that achieve the best performance on the validation set to compare. 
In Caser, the number of horizon convolution kernels is set as 64, the number of vertical convolution kernels is set as 32 and the union window size is set as 5. In GRU4Rec, the dimension of the hidden layer is set as 64. In SARS and TiSARS, the number of heads is set as 1. In SLRS+ and Chorus, the learning rate is set as 5e-4. In AHMP, both the head number and layer number are set as 1. The learning rate is set as 1e-4. In KDA, the number of heads is set as 4, and the learning rate is set as 1e-3. Besides, the learning of all the models is carried out five times to report the average results.

\end{document}